\documentclass[onecolumn,preprintnumbers,amsmath,amssymb]{revtex4}

\begin{document}


\title{Deformed dispersion relations and the degree of coherence function}
\author{Abel Camacho}
\email{acq@xanum.uam.mx} \affiliation{Departamento de F\'{\i}sica,
Universidad Aut\'onoma Metropolitana--Iztapalapa\\
Apartado Postal 55--534, C.P. 09340, M\'exico, D.F., M\'exico.}

\author{Alfredo Mac\'{\i}as}
\email{amac@xanum.uam.mx} \affiliation{Departamento de
F\'{\i}sica,
Universidad Aut\'onoma Metropolitana--Iztapalapa\\
Apartado Postal 55--534, C.P. 09340, M\'exico, D.F., M\'exico.}

\date{\today}

\begin{abstract}
The analysis of the modifications that the presence of a deformed
dispersion relation entails in the roots of the so--called degree
of coherence function, for a beam embodying two different
frequencies and moving in a Michelson interferometer, is carried
out. The conditions to be satisfied, in order to detect this kind
of quantum gravity effect, are also obtained.
\end{abstract}
\maketitle

\section{Introduction}

Amid the gamut of approaches, whose goal is a quantum theory of
gravity, some clearly protrude, namely, they entail the
modification of the dispersion relation \cite{[1]}. These ideas
appear in several models, for instance, quantum--gravity
approaches based upon non--commutative geometry \cite{[2], [3]},
or loop--quantum gravity models \cite{[4], [5]}, etc. In them
Lorentz symmetry becomes only an approximation for quantum space
\cite{[6], [7], [8]}, and do entail modifications in some
fundamental physical concepts, as the uncertainty principle
\cite{[9]}, for instance.

The quest for quantum--gravity effects is not restricted to the
case of dispersion relations. Indeed, Dirac equation can also be
employed \cite{[10]} in this context, and hence we may look for
the consequences of this sort of models in the motion equation of
1/2 spin particles. For instance, the spreading of a wave packet
is modified, and in principle, it is possible to detect effects
induced by quantum--gravity theory by monitoring this parameter,
or, as already shown \cite{[10]}, Larmor precession presents a
novel dependence, and in consequence the angular velocity of the
expectation values of the components of the spin allows us to test
our modified Dirac equation.

In the present work the consequences, upon the roots of the
so--called degree of coherence function \cite{[11]}, of a deformed
dispersion relation are analyzed. This will be done for the case
of a beam comprising two di\-fferent frequencies. At this point it
is noteworthy to comment the existence, already, of a work
\cite{[12]} containing a qualitative analysis of the modifications
emerging in the interference pattern of a Michelson device. In the
aforementioned work the modifications in the phase shifts have
been studied, nevertheless the changes in the roots of the degree
of coherence function has not yet been addressed. The conditions
to be sa\-tisfied, in order to detect this kind of quantum gravity
effect, will be also deduced.

Finally, some words will be said concerning the possible future
work in the realm of deformed dispersion relations and its
detection resorting to higher--order coherence effects (such as
the so--called Hanbury--Brown--Twiss effect \cite{[11]}), or to
non--monocromatic light sources, a case that, by the way, up to
now strangely has not been seriously considered, and the one could
shed some light upon the present issue.
\bigskip
\bigskip

\section{Degree of coherence function and deformed dispersion relations}
\bigskip
\bigskip

As already mentioned above several quantum--gravity models predict
a modified dispersion relation \cite{[1],[2], [3], [4], [5]}, the
one can be characterized, phenomenologically, through corrections
hinging upon Planck's length, $l_p$,

\begin{equation}
E^2 = p^2\Bigl[1 - \alpha\Bigl(El_p\Bigr)^n\Bigr].
\label{Disprel1}
\end{equation}

Here $\alpha$ is a coefficient, usually of order 1, and whose
precise value depends upon the considered quantum--gravity model,
while $n$, the lowest power in Planck's length leading to a
non--vanishing contribution, is also model dependent. Casting
(\ref{Disprel1}) in ordinary units we have

\begin{equation}
E^2 = p^2c^2\Bigl[1 -
\alpha\Bigl(E\sqrt{G/(c^5\hbar)}\Bigr)^n\Bigr]. \label{Disprel2}
\end{equation}

The expression

\begin{equation}
p =\hbar k, \label{Mom1}
\end{equation}

leads us to

\begin{equation}
k =\frac{E/(c\hbar)}{\Bigl[1 -
\alpha\Bigl(E\sqrt{G/(c^5\hbar)}\Bigr)^n\Bigr]^{1/2}}. \label{k1}
\end{equation}

Since we expect very tiny corrections, then the following
expansion is justified

\begin{equation}
k =\frac{E}{c\hbar}\Bigl[1 +
\frac{\alpha}{2}\Bigl(E\sqrt{G/(c^5\hbar)}\Bigr)^n +
\frac{3}{8}\alpha^2\Bigl(E\sqrt{G/(c^5\hbar)}\Bigr)^{2n}+...\Bigr].
\label{k2}
\end{equation}

Let us now consider two beams with energies $E_1$ and $E_2$,
respectively, such that $E_2 = E_1 + \Delta E$, and in addition,
it will be assumed that they interfere in a Michelson device
\cite{[11]}. As is already known each frequency produces an
interference pattern, and at this point it will be su\-pposed that
the corresponding beat frequency is to high to be detected
\cite{[11]}, i.e., the output intensity is obtained adding the
intensities associated with each frequency contained in the input.
Under these conditions the measured intensity reads

\begin{equation}
I = I_1\Bigl[1 + \cos(\omega_1\tau_1)\Bigr] + I_2\Bigl[1 +
\cos(\omega_2\tau_2)\Bigr]. \label{Intensity1}
\end{equation}

In this last expression $I_1$ and $I_2$ denote the intensities of
the two beams, $\omega_1$, $\omega_2$ the corresponding
frequencies, and

\begin{equation}
\tau_1 =2d/c_1, ~~\tau_2 =2d/c_2. \label{Time}
\end{equation}
Here $d$ is the difference in length in the two interfero\-meter
arms, and $c_1$, $c_2$, the corresponding velocities, here the
velocity has a non--trivial energy dependence \cite{[1]}, i.e.,
$c_1 \not=c_2$.

From now on we will assume that $I_1 = I_2$, such that $I_0=I_1+
I_2$, therefore the detected intensity can be cast in the
following form

\begin{equation}
I = I_0\Bigl[1 + \gamma(d)\Bigr]. \label{Intensity2}
\end{equation}
In this last equation the so--called degree of coherence function
$\gamma(d)$ has been introduced \cite{[11]}, the one for our
situation reads ($k_1$ and $k_2$ are the corresponding wave
numbers)

\begin{equation}
\gamma(d)= \cos\Bigl([k_1 + k_2]d/2\Bigr)\cos\Bigl([k_1 -
k_2]d/2\Bigr). \label{Degree1}
\end{equation}

A fleeting glimpse at (\ref{k1}) clearly shows that
(\ref{Degree1}) does depend upon $\alpha$ and $n$, and in
consequence the roots of the degree of coherence function will be
modified by the presence of a deformed dispersion relation.

The expression providing us the roots  of the degree of coherence
function is

\begin{equation}
\Bigl(k_1 - k_2\Bigr)d/2 = \pi/2. \label{Root1}
\end{equation}

Resorting to our previous expressions we may rewrite (\ref{Root1})
as

\begin{eqnarray}
d = c\hbar\pi\Bigl\{\Delta E +
\frac{\alpha}{2}E_1\Bigl(E_1\sqrt{G/(c^5\hbar)}\Bigr)^n\nonumber\\
\times\Bigl[(n+1)\frac{\Delta E}{E_1}
+\frac{n(n+1)}{2}\Bigl(\frac{\Delta E}{E_1}\Bigr)^2 +...
\Bigr]\Bigr\}^{-1}. \label{Root12}
\end{eqnarray}

Let us now define $\beta = \Delta E/E_1$, a real number smaller
than 1. In the present proposal we will consider two possible
values for $n$, namely:
\bigskip
\bigskip

\subsection{Case n= 1}
\bigskip
\bigskip

For this situation we have that the roots of the degree of
coherence function become, approximately

\begin{eqnarray}
d = \frac{c\hbar\pi}{E_1}\Bigl\{\beta
-\frac{\alpha}{2}\Bigl(E_1\sqrt{G/(c^5\hbar)}\Bigr) \Bigl[2 +
\beta\Bigr]\Bigr\}. \label{Root3}
\end{eqnarray}

For the sake of clarity let us assume that $\alpha\sim 1$, a
restriction that is not devoid of physical content \cite{[1]}. The
possibility of detecting this deformed dispersion relation will
hinge upon the fulfillment of the condition

\begin{eqnarray}
\vert D - d\vert > \Delta d. \label{Exp1}
\end{eqnarray}
In this last equation $D$ denotes the usual value in the
difference of the interferometer arms at which the degree of
coherence function vanishes (that is when $\alpha =0$), whereas
$\Delta d$ is the corresponding experimental resolution. This can
be cast in the following form

\begin{eqnarray}
\frac{\Delta E}{E_1}> \frac{2\Delta d}{\pi l_p} - 1. \label{Exp2}
\end{eqnarray}

 Recalling that from square one it was assumed that our device
 cannot detect the beat frequencies, i.e., if $T$ denotes the time
 resolution of the measuring device, then

\begin{eqnarray}
\vert\omega_2 - \omega_1\vert T/2>>1. \label{Exp3}
\end{eqnarray}
This last condition may be rewritten as

\begin{eqnarray}
T\Delta E>\hbar. \label{Exp4}
\end{eqnarray}

In other words, (\ref{Exp2}) and (\ref{Exp4}) are the two
conditions to be fulfilled if the case $n=1$ and $\alpha \sim 1$
is to be detected.
\bigskip
\bigskip

\subsection{Case n= 2}
\bigskip
\bigskip

Under these conditions ($n = 2$ and $\alpha \sim 1$) the roots of
the degree of coherence function read, approximately

\begin{eqnarray}
d = \frac{c\hbar\pi}{E_1}\Bigl\{\beta
-\frac{\alpha}{2}\Bigl(E_1\sqrt{G/(c^5\hbar)}\Bigr)^2\Bigl[3 +
3\beta+ \beta^2\Bigr]\Bigr\}. \label{Root4}
\end{eqnarray}

The expression tantamount to (\ref{Exp2}) is

\begin{eqnarray}
\Bigl\{3 + 3\Bigl(\frac{\Delta E}{E_1}\Bigr) + \Bigl(\frac{\Delta
E}{E_1}\Bigr)^2\Bigr\}E_1> 2\frac{c\hbar\Delta d}{\pi l^2_p}.
\label{Exp5}
\end{eqnarray}

The impossibility of detecting beat frequencies translates, once
again, as

\begin{eqnarray}
T\Delta E>\hbar. \label{Exp6}
\end{eqnarray}
\bigskip
\bigskip

\section{Conclusions}
\bigskip
\bigskip

In the present work the possibility of detecting two different
deformed dispersion relations, resorting to the analysis of the
roots of the degree of coherence function, has been carried out.

The impossibility of detecting beat frequencies renders only one
condition, see expressions (\ref{Exp4}) and (\ref{Exp6}). The
experimental difficulty appears in connection with (\ref{Exp2})
and (\ref{Exp5}), which are the restrictions to be satisfied in
order to detect this kind of effects. Forsooth, since we have,
from square one, imposed the condition $\frac{\Delta E}{E_1} <1$,
then (\ref{Exp2}) entails a very stringent restriction, namely a
experimental resolution very close to Planck's length, i.e.,
$\Delta d\sim l_p$.

The case $n=2$ becomes even worse, as usual \cite{[1]}.  A rough
estimate of the required energy, for the case in which $\Delta
d\sim 10^{-4}$cm, renders energies higher than the so--called GZK
limit for cosmic rays \cite{[13]}. A fleeting glimpse at
(\ref{Exp5}) clearly shows us that in this case the problem stems
from the presence of the factor $l^2_p$.

The conclusions that have been drawn from the previous analysis
have a not very optimistic atmosphere, and in consequence we may
wonder if interferometry could be useful in the present context.
In order to address this issue let us recall that in the extant
literature the considered experimental proposals do reduce to the
case of first order coherence experiments \cite{[12]} (Michelson
interferometer falls within this category \cite{[11]}), or they
consider only a finite number of monocromatic sources \cite{[12]}.
At this point it is noteworthy to comment that optical experiments
offer a much richer realm of possibilities. Forsooth,
higher--order coherence effects, for instance, the so called
Hanbury--Brown--Twiss effect \cite{[11]}), could be also
considered and explored within the present context, or the effects
of a deformed dispersion relation upon a light source with a
continuous frequency distribution could be studied. The results of
the analysis of the aforementioned proposals will be published
elsewhere.

\begin{acknowledgments}
We dedicate the present work to Michael Ryan on occa\-sion of his
$60^{th}$ birthday. This research was supported by CONACYT Grant
42191--F. A. C. would like to thank A.A. Cuevas--Sosa for useful
discussions and literature hints.
\end{acknowledgments}


\begin{thebibliography}{}

\bibitem{[1]} G. Amelino--Camelia, Int. J. Mod. Phys. {\bf D9}, 1633 (2003).

\bibitem{[2]} G. Amelino--Camelia and T. Piran, Phys. Rev. {\bf D64}, 036005 (2001).

\bibitem{[3]} G. Amelino--Camelia, Nature {\bf 418}, 661 (2000).

\bibitem{[4]} R.Gambini and J. Pullin, Phys. Rev. {\bf D59}, 124021 (1999).

\bibitem{[5]} L.Smolin, LANL hep--th/0303185.

\bibitem{[6]} G. Amelino--Camelia, Nature {\bf 398}, 216 (1999).

\bibitem{[7]} G. Amelino--Camelia, J. Ellis, N. E. Mavromatos, D.
V. Nanopoulos, and S. Sarkar, Nature {\bf 393}, 763 (1998).

\bibitem{[8]} J. Alfaro, H. A. Morales--Tecotl, and L. F. Urrutia, Phys.
Rev. Lett. {\bf 84}, 2318 (2000).

\bibitem{[9]} A. Camacho, Gen. Rel. Grav. {\bf 34}, 1839 (2002).

\bibitem{[10]} A. Camacho and  A. Macias, Phys. Lett. {\bf B582}, 229 (2004).

\bibitem{[11]}  L. Mandel and E. Wolf, {\em Optical Coherence and Quantum Optics},
Cambridge University Press, Cambridge (1995).

\bibitem{[12]} G. Amelino--Camelia and C. L\"ammerzahl, Class.
Quantum Grav.{\bf 21}, 899--915 (2004).

\bibitem{[13]} T. Piran, {\em Gamma--Ray Bursts as Probes for Quantum Gravity},
arXiv: astro--ph/0407462.


\end{thebibliography}
\end{document}